# From Prediction to Action: Critical Role of Performance Estimation for Machine-Learning-Driven Materials Discovery

**Mario Boley**
Monash University, Department of Data Science and AI
mario.boley@monash.edu

**Felix Luong**
Monash University, Department of Data Science and AI
felix.luong@monash.edu

**Simon Teshuva**
Monash University, Department of Data Science and AI
simon.teshuva@monash.edu

**Daniel F. Schmidt**
Monash University, Department of Data Science and AI
daniel.schmidt@monash.edu

**Lucas Foppa**
NOMAD Laboratory at the Fritz Haber Institute of the Max-Planck-Gesellschaft and IRIS-Adlershof of the Humboldt-Universität zu Berlin
foppa@fhi-berlin.mpg.de

**Matthias Scheffler**
NOMAD Laboratory at the Fritz Haber Institute of the Max-Planck-Gesellschaft and IRIS-Adlershof of the Humboldt-Universität zu Berlin
scheffler@fhi-berlin.mpg.de

***Abstract*** *Materials discovery driven by statistical property models is an iterative decision process, during which an initial data collection is extended with new data proposed by a model-informed acquisition function—with the goal to maximize a certain "reward" over time, such as the maximum property value discovered so far. While the materials science community achieved much progress in developing property models that predict well on average with respect to the training distribution, this form of in-distribution performance measurement is not directly coupled with the discovery reward. This is because an iterative discovery process has a shifting reward distribution that is over-proportionally determined by the model performance for exceptional materials. We demonstrate this problem using the example of bulk modulus maximization among double perovskite oxides. We find that the in-distribution predictive performance suggests random forests as superior to Gaussian process regression, while the results are inverse in terms of the discovery rewards. We argue that the lack of proper performance estimation methods from pre-computed data collections is a fundamental problem for improving data-driven materials discovery, and we propose a novel such estimator that, in contrast to naïve reward estimation, successfully predicts Gaussian processes with the "expected improvement" acquisition function as the best out of four options in our demonstrational study for double perovskites. Importantly, it does so without requiring the over thousand ab initio computations that were needed to confirm this prediction.*

**Status**

In recent years, the materials science community has established a large-scale infrastructure for data sharing that promises to increase the efficiency of the "data-driven" discovery of novel useful materials [1]. Growing data collections are envisioned to lead to increasingly accurate statistical models for property prediction that can significantly reduce the number of necessary experiments or first principles computations and, thus, substantially improve the cost and time for critical discoveries [2]. Indeed, the combination of public datasets and robust statistical estimation techniques like cross validation (CV) enables a collaborative improvement process ("common task framework" [3, 4]). As a result, there are now models that can predict certain materials properties well *on average* with respect to the same distribution as the training data. Unfortunately, the *in-distribution* expected performance, as estimated by CV, is not directly coupled with the performance for the discovery of novel materials: expected performance fails to capture the model behavior for the very few exceptional materials that one aims to discover, and, fundamentally, in-distribution performance is irrelevant for a discovery process that is designed to generate high-performing materials more frequently than they occur in the initial training data.

Recognizing these issues, the community increasingly focusses on active learning approaches [5] like Bayesian optimization for model-driven blackbox optimization [6] (BBO). These methods manage an iterative modelling and data acquisition process and aim to optimize the cumulative "reward" received for the acquired data points over time, such as the maximum property value discovered so far. This process, illustrated in Figure 1, is enabled by an acquisition function that leverages the predictions of a statistical model together with its uncertainty quantification to effectively manage the underlying trade-off of exploration (learning more about the candidate space) and exploitation (aim to sample high value candidates). This shift to consider actions instead of just predictions constitutes an important step towards accelerated materials discovery, but it reveals shortcomings not only in existing modelling approaches but more fundamentally in the methodological framework used to improve those models. In particular, the inapplicability of established performance estimation frameworks based on pre-generated data renders it extremely costly to conclusively compare and to systematically improve methods.

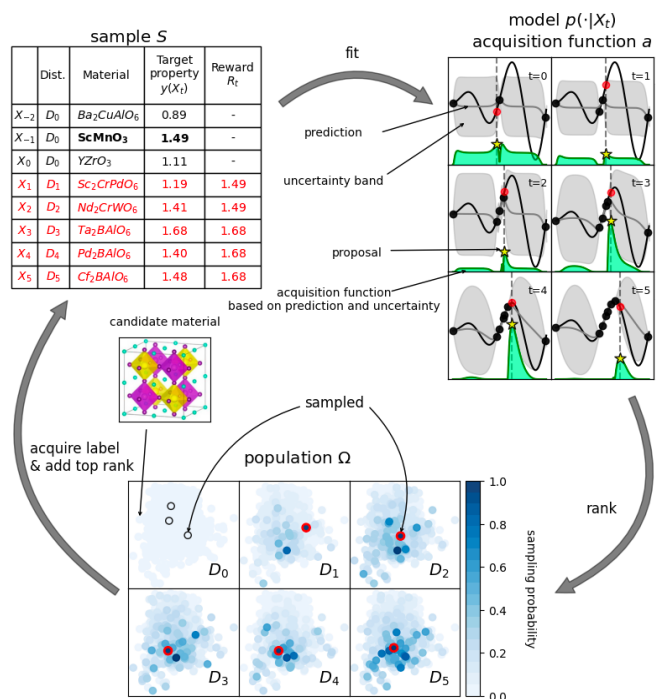

*Figure 1. Schematic steps of iterative model-driven discovery process. At time $t$: (i) probabilistic property model is fitted to sample $\{X_{-N+1}, \ldots, X_0; X_1; \ldots; X_{t-1}\}$ of materials population $\Omega$, i.e., a conditional density function $p(y \mid x)$ is learned that provides probability density of property value $y$ for material $x$, which gives rise to both (mean) prediction $f(x) = \mathbb{E}_p(Y \mid X = x)$ and uncertainty (variance) $\sigma^2(x) = \mathbb{V}_p(Y \mid X = x)$ where expected value and variance are taken with respect to $p$; (ii)*

*remaining population is ranked by acquisition function, e.g., "expected improvement" of reward $a(x) = \mathbb{E}_p(R_t - R_{t-1} | X_t = x)$, which for conditionally normal property models can be computed as $a(x) = f(x) + \sigma^2(x)p(R_t | x)/(1 - P(R_t | x))$ where $P$ is the modelled cumulative distribution function; and (iii) label for top-ranked material is acquired and added to data sample generating reward, e.g., defined as $R_t = \max\{y(X_i): -N < i \leq t\}$ when maximizing a single property or figure of merit $y$, which incentivizes the discovery of materials with high $y$-value as early as possible in the process. While standard statistical analysis assumes the initial data points $X_{-N+1}, \ldots, X_0$ to be drawn with respect to some sampling distribution $D_0$, this distribution does not have to be balanced or representative of the whole population. However, any concentration away from a representative, i.e., uniform, sampling distribution, poses the risk of delayed reward generation, and a misspecified acquisition function or model, in particular one with over-confident predictions, even risks to never escape local maxima represented in the initial data collection. The sampling distribution of subsequent points $D_1, D_2, \ldots, D_T$ vary and depend on the combination of model $p$ and acquisition function $a$. Hence, they cannot be pre-generated for new methods rendering label generation a key bottleneck in method development.*

**Current and Future Challenges**

To illustrate these challenges, let us consider as example the discovery of double perovskite oxides with high *ab initio* computed bulk modulus, where we use two popular statistical models, Gaussian process (GP) regression and random forest (RF), and two BBO data acquisition strategies, *expected improvement* [7] (EI) of rewards and *pure exploitation* [8] (XT). GPs are the traditional BBO model, because their Bayesian approach provides a principled quantification of "epistemic" uncertainty, i.e., uncertainty from a lack of training data related to a specific test point. However, they can struggle already with moderately high-dimensional representations such as the 24 features used in this example. In contrast, RFs are known to work robustly well with high-dimensional feature spaces [9], while their ensemble-based uncertainty quantification does not represent epistemic uncertainty. Interestingly, as shown in Figure 2, CV indicates that RF has the better in-distribution predictive performance not only in terms of squared error but also in terms of log loss, which takes uncertainty into account. Nevertheless, RF is outperformed by GP in terms of the produced discovery rewards, demonstrating that standard in-distribution performance estimation techniques can suggest sub-optimal methods.

This demonstrates that already method selection is a real challenge for practical problems. However, the situation is much worse for methodological research that aims to not only determine, which of a small number of established methods works best, but to test dozens of combinations of models and acquisition functions. Absent innovation in performance estimation, comparing $K$ methods in terms of their expected discovery reward across $L$ repetitons of $T$ rounds requires the acquisition of $KLT$ labels in addition to any pre-generated initial data. This is because, even when starting from a common initial training distribution, each method produces its own sequence of proposal distributions. Since these distributions are unknown *a priori*, there is no way to pre-generate data from them, blocking the usual collaborative improvement process around an initially released dataset. Thus, the prohibitive cost of expected reward estimation currently blocks substantial progress in addressing other important challenges like unsound uncertainty quantification or acquisition function optimization with infinite candidate populations particularly when using non-invertible materials representations.

**Advances in Science and Technology to Meet Challenges**

Given these considerations, a central research goal should be to find reliable approaches for estimating a method's expected discovery reward based on existing data. A simple but infeasible state-of-the-art strategy is to run a method repeatedly using sub-samples of size $n$ from the given dataset as initial data and the sub-sample complement as candidate pool, such that the ratio $n/N$ is close to $N/M$ where $M$ is the overall population size. That is, one naively uses the initial dataset as proxy for the population. For at least two reasons, this simplistic approach is likely to produce misleading results (see Figure 2, middle left). Firstly, the real rewards are determined by the exceptional materials in the tail of the target property distribution, which are almost certainly not well represented in the available dataset. Secondly, changing the absolute sizes of initial data and candidate population misestimates model performance and, more severely, misrepresents the real overwhelming number of uninteresting materials that an efficient search must largely avoid.

Here, we present an adjusted reward estimation approach that provides random initial and candidate sets with realistic absolute numbers of unrepresented exceptional materials as well as distinct ordinary materials to distract from them. Let $X_{(1)}, \ldots, X_{(N)}$ denote the initial data elements in increasing order of their target property or figure of merit values. Based on an estimate $\hat{\alpha}$ of the unrepresented fraction of top materials $\alpha = \#\{X \in \Omega: y(X) \geq y(X_{(N)})\}/M$ create:

1. **an initial dataset** by drawing a size-$N$ bootstrap sub-sample [10], i.e., sample with replacement, from the low property value materials $X_{(1)}, \ldots, X_{(N-\lceil \hat{\alpha} M \rceil - 1)}$ and
2. **a candidate set** consisting of the held-out top $\lceil \hat{\alpha} M \rceil$ materials and an up-sampled and stochastically perturbed set $\tilde{X}_1, \ldots, \tilde{X}_{M-\lceil \hat{\alpha} M \rceil}$ from the unsampled elements of the bootstrap sample.

As shown in Figure 2 (bottom left), reward estimation with this approach performs much better than naïve estimation for our bulk modulus example. It accurately predicts GP with EI to produce the highest bulk modulus and highest cumulative reward out of the four candidate methods. As desired, this is based entirely on the initially available data without requiring the over thousand additional calculations that were needed to confirm this result.

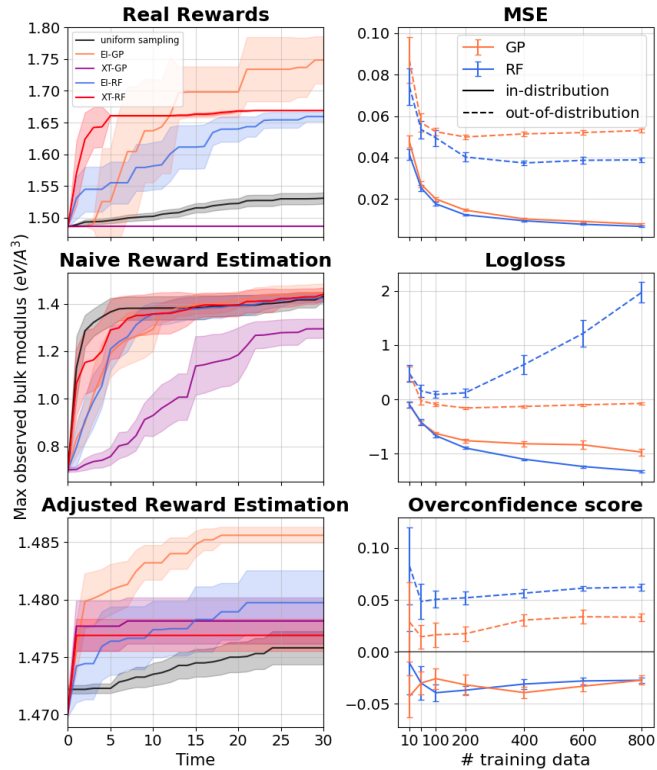

*Figure 2. Performance of Gaussian process (GP) and random forest (RF) models for discovering double perovskites with high bulk modulus.* **Left column:** *Rewards generated by models with either expected improvement (EI) or pure exploitation (XT) acquisition function as well as their naïve and adjusted reward estimation. Real rewards are mean rewards based on five (EI-GP), nine (EI-RF), ten (both XT), or 100 (uniform) simulations. Estimated rewards are the mean of 20 sub-sampling repetitions. All error bars correspond to 90% confidence intervals. GP with EI has the highest mean reward (1.66 $eV/A^3$) and discovers the highest bulk modulus (1.75 $eV/A^3$ on average) in 35 rounds, which is qualitatively predicted by adjusted reward estimation.* **Right column:** *Model predictive performance estimates in terms of the mean squared error MSE $\mathbb{E}_D(y(X) - f(X))^2$ where $f(X)$ is the prediction for random input point $X$ with property value $y(X)$, log loss $\mathbb{E}_D(\log p(f(X) | X))$ where $p$ is the modelled density of $y(X)$, and overconfidence score $\mathbb{E}(|y(X) - f(X)| - \sigma(X))$ where $\sigma(X)$ where is the modelled standard deviation of $y(X)$ given $X$. Here, all expected values refer to unknown true distributions estimated via 10 (over-confidence score) and 20 (MSE and log loss) repetitions of sub-sampling with replacement from*

*available data (i.e., bootstrap sampling). In-distribution performance is performance with respect to the initial sampling distribution $D_0$, out-of-distribution is with respect to the uniform mixture of the distributions $D_1$ to $D_{100}$ of the data points examined by the discovery process. While RF provides a better mean squared error, both in- and out-of-distribution, its out-of-distribution log loss is increasing with the size of the training, indicating a failure of its uncertainty quantification.*

**Concluding Remarks**

The lack of reliable approaches to estimate expected discovery rewards from a given dataset is a serious roadblock for the development of active learning methods for materials discovery. Without such estimators, the evaluation of each candidate method requires the acquisition of a potentially large number of labels in addition to any initially available data collection, preventing the usual collaborative process that led to fast-paced improvements of predictive model performance with fixed distributions.

Naïve reward estimation from the initial data typically fails because of unsuitable data proportions and underrepresented extreme events. We presented an adjusted approach that, by correcting for these factors, successfully assesses which combination of acquisition function and statistical model works best for the exemplary task of double perovskite bulk modulus optimization. This or similar approaches could become efficiently computable proxies for real method performances and thus enable fast community-driven improvements to data-driven methods for materials discovery.

**Acknowledgements**
*This work was supported by the Australian Research Council (DP210100045) and the ERC Advanced Grant TEC1p (European Research Council, Grant Agreement No. 740233).*

**References**

[1] M. Scheffler, M. Aeschlimann, M. Albrecht, T. Bereau, H.-J. Bungartz, C. Felser, M. Greiner, A. Groß, C. T. Koch, K. Kremer and others, "FAIR data enabling new horizons for materials research," *Nature,* vol. 604, no. 7907, pp. 635-642, 2022.

[2] J. Schmidt, M. R. Marques, S. Botti and M. A. Marques, "Recent advances and applications of machine learning in solid-state materials science," *npj Computational Materials,* vol. 5, no. 1, p. 83, 2019.

[3] D. Donaho, "50 Years of Data Science," *Journal of Computational and Graphical Statistics,* vol. 26, no. 4, pp. 745-766, 2017.

[4] C. Sutton, L. M. Ghiringhelli, T. Yamamoto, Y. Lysogorskiy, L. Blumenthal, T. Hammerschmidt, J. R. Golebiowski, X. Liu, A. Ziletti and M. Scheffler, "Crowd-sourcing materials-science challenges with the NOMAD 2018 Kaggle competition," *npj Computational Materials,* vol. 5, no. 1, p. 111, 2019.

[5] T. Lookman, P. V. Balachandran, D. Xue and R. Yuan, "Active learning in materials science with emphasis on adaptive sampling using uncertainties for targeted design," *npj Computational Materials,* vol. 5, no. 1, p. 21, 2019.

[6] B. Shahriari, K. Swersky, Z. Wang, R. P. Adams and N. de Freitas, "Taking the Human Out of the Loop: A Review of Bayesian Optimization," *Proceedings of the IEEE,* vol. 104, no. 1, pp. 148--175, 2016.

[7] D. Zhan and H. Xing, "Expected improvement for expensive optimization: a review," *Journal of Global Optimization,* vol. 78, no. 3, pp. 507-544, 2020.

[8] G. De Ath, R. M. Everson, A. A. Rahat and J. E. Fieldsend, "Greed is good: Exploration and exploitation trade-offs in Bayesian optimisation," *ACM Transactions on Evolutionary Learning and Optimization,* vol. 1, no. 1, pp. 1-27, 2021.

[9] G. Biau and E. Scornet, "A random forest guided tour," *Test,* vol. 25, pp. 197--227, 2016.

[10] E. B, "Bootstrap Methods: Another Look at the Jackknife," *Ann. Statist.,* vol. 7, no. 1, pp. 1-26, 1979.